\documentclass[aps,pra,twocolumn,superscriptaddress,showpacs]{revtex4}   
\usepackage{amsmath,amssymb,amstext,graphicx,bm,bbm}
\usepackage{psfrag}
\makeatletter

\begin{document}
\title{Holonomic Quantum Computation with Electron Spins in Quantum Dots}

\author{Vitaly N. Golovach}
\affiliation{Arnold Sommerfeld Center for Theoretical Physics and Center for Nanoscience
Department of Physics,
Ludwig-Maximilians-Universit\"at,
Theresienstrasse 37,
D-80333 Munich,
Germany}

\author{Massoud Borhani}
\affiliation{Department of Physics, 
University of Basel, Klingelbergstrasse 82, 
4056 Basel, Switzerland}
\affiliation{Department of Physics, University at Buffalo,
 SUNY, Buffalo, NY 14260-1500, USA}

\author{Daniel Loss}
\affiliation{Department of Physics, 
University of Basel, Klingelbergstrasse 82, 
4056 Basel, Switzerland}

\date{\today}

\begin{abstract}
With the help of the spin-orbit interaction, we propose a scheme to perform 
holonomic single qubit gates on the electron spin confined to a quantum dot. 
The manipulation is done in the 
absence (or presence) of an applied magnetic field. By adiabatic changing the position 
of the confinement potential, one can rotate the spin state of the electron 
around the Bloch sphere in semiconductor heterostructures. The dynamics of the system 
is equivalent to employing an effective non-Abelian gauge potential 
whose structure depends on the type of the spin-orbit interaction.
As an example, we find an analytic expression for the electron spin dynamics
when the dot is moved around a circular path
 (with radius $R$) on the two dimensional electron gas (2DEG), and show that all
single qubit gates can be realized by tuning the radius and orientation of the circular paths.
Moreover, using the Heisenberg exchange interaction, we demonstrate how one can generate two-qubit gates
by bringing two quantum dots near each other, yielding a scalable scheme to perform quantum
computing on arbitrary $N$ qubits.
 This proposal shows a way of realizing holonomic quantum 
computers in solid-state systems.    
\end{abstract}

\maketitle

\section{Introduction}
The emergence of geometrical phases in quantum mechanical systems and their  physical and
 geometrical consequences were first recognized by Berry and Simon in their works on cyclic quantum evolution \cite {Berry,Simon}.
 Soon after that, Wilczek and Zee discovered the connection
 between (non-Abelian) gauge fields and the adiabatic dynamics of such systems, 
where the dimension of non-Abelian geometric phases is given by the n-fold degeneracy 
of the eigenstates of the Hamiltonian \cite {WZ}. 
Moreover, Aharonov and Anandan \cite{Ah-An} generalized Berry's idea to non-adiabatic
evolutions, however, the resulting geometric phase is not given anymore by the holonomy in the
parameter space of the Hamiltonian but in the projective Hilbert space.
Although, at the time, geometrical phases were already known in classical systems 
\cite {WSh},
their quantum mechanical counterparts are physically richer and more subtle.

Based on the above mentioned ideas, a variety of schemes for holonomic (HQC) and geometric (GQC) 
quantum computation have been proposed, which recently attained a considerable attention, 
and believed to be promising candidates to implement
quantum computers using topological transformations as qubit gates 
\cite{Zanardi,Erik}. 
In HQC, for example, one can perform quantum computing
 by encoding quantum information in the degenerate levels of the 
Hamiltonian and adiabatically traversing closed loops (holonomies) in the parameter
 space of the Hamiltonian.
So far, many theoretical (and experimental) investigations have been made to implement such Hamiltonians in physical systems, for example,  confined ions in a linear Pauli trap \cite{Zoller}. The experiment by Jones {\it et al.} \cite {Jones} was the first attempt in this direction where they realized geometric two-qubit gates between a pair of nuclear spins. Note that geometrical phases are generally small compared to dynamical phases. Being a small effect on top of a large effect, makes it challenging for the experimentalists to identify and employ them for quantum computation.

Among several proposals for HQC and GQC, solid state matrix 
is usually more desirable due to its potential in realizing large scale qubit systems.
Specifically, spin of an electron in a quantum dot, as a two level system, has been shown to be a suitable 
qubit \cite{LD}, meanwhile, rapid experimental progress in the field of semiconductor
 spintronics made it possible to access individual electron spin in
 low dimensional systems \cite{ALS}.
Manipulating the spin of the electrons/holes in semiconductors is, therefore,
 one of the objectives of spintronics \cite{ALS,BV-Review}.
Among different tools to achieve this goal is to apply an external magnetic field, in combination
with the spin-orbit interaction, in a controlled way.
Recently, there has been a great progress in developing
 techniques to manipulate electrically 
the electron/hole spins in two dimensional electron/hole gases (2DEGs/2DHGs) and quantum dots (QDs)
\cite{RashbaEfrosPRL, RashbaEfrosAPL,MathiasEDSR,GBL,Flindt,Nitta,Salis,KatoScience,
KatoNature,Denis}.
However, most of the previous works on confined electrons are based on the assumption that the 
quantum dot itself is almost frozen in real space. 
Moreover, the presence of an applied magnetic field is usually assumed in order to break 
the time reversal symmetry, which is essential in electron spin resonance (ESR) schemes. The question is: what is the dynamics
of the spin sector if we move the quantum dot in the absence 
(or presence) of the magnetic field?
If time reversal symmetry is not broken, a convenient way to study the dynamics of such a 
system is to employ {\it non-Abelian} gauge fields \cite {WZ}. 
In spite of the fact that the origin of non-Abelian gauge fields in classical/quantum field theories 
and the current problem is somewhat different, as long as the dynamics is concerned, they
play the same role.

Using an effective non-Abelian gauge potential \cite{WZ,BV-Review,Pablo}, 
we propose a novel technique to manipulate topologically the spin of an electron
inside a quantum dot without using any {\it applied} magnetic field.
Although the rotation of the the electron spin in a moving quantum dot has been 
studied in previous works \cite{BV-Review,Pablo},
our goal here is to implement systematically {\it {all}} necessary single qubit gates
for quantum information processing.
We consider a setup, where the quantum dot can be moved on the
substrate at distances comparable to $\lambda_{SO}$, the spin-orbit length.
In addition, we assume that the electron is strongly confined to the quantum dot
at a length scale $\lambda_d \ll \lambda_{SO}$, where $\lambda_d$ is the dot size .
We study the case which the confining potential is only displaced parallel to itself 
by a vector ${\bm r}_0(t)$ without changing its shape (see Fig. \ref{trajectoryfigure}).
Moreover, we assume that the driving electric field is classical and only quantize the 
electron dynamics (for the discussion of the quantum fluctuations of the electromagnetic fields which lead to
the decoherence of the Kramers doublets, see Ref. \cite{Pablo}).
 For a moving quantum dot around a circular path,
we derive an exact solution of the Schr\"odinger equation for the spin sector,
 in the first order in spin-orbit interaction.
In addition, we show how one can generate different single- and two-qubit 
rotations and perform quantum computing on $N$ spins.

\section{The Model and Basic Relations}
\begin{figure}
\begin{center}
\includegraphics[angle=0,width=0.4\textwidth]{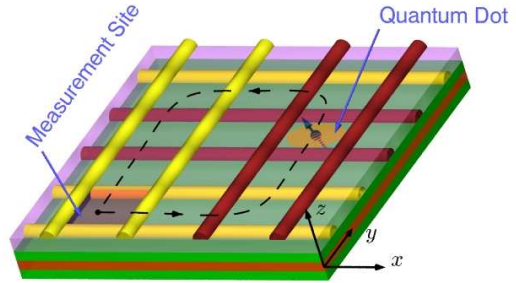}
\caption{\small
(Color online)
A set of metallic gates, deposited on top of the 
heterostructure, control the quantum dot position
in the ($x$,$y$)-plane. The gates that define the current
dot position are highlighted in a darker color.
The superimposed gates are separated from each other 
by an insulating layer.
The measurement site is used to initialize and read out the
spin state of the quantum dot with the help of additional
controls (not shown here). 
By applying time-dependent voltages to the 
gates, one is able to move the quantum dot along a desired 
trajectory, thus forming, for instance, a Wilson loop (dashed line).
}
\label{Setup1}
\end{center}
\end{figure}

\begin{figure}
\begin{center}
\includegraphics[angle=0,width=0.4\textwidth]{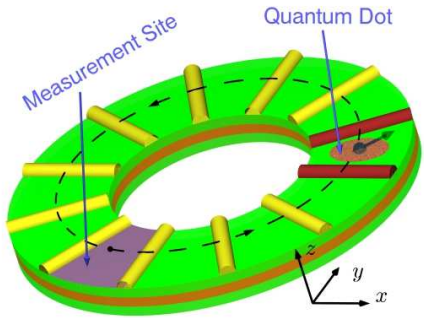}
\caption{\small
(Color online)
Same as in Fig.~\ref{Setup1}, except that the
quantum dot is defined using a single layer
of metallic gates deposited on top of a
ring. 
The ring can be obtained out of a heterostructure
by means of etching or defined 
with the help of an atomic force microscope
using the oxidation technique of Refs.~[\onlinecite{HeldEnsslin},\onlinecite{FuhrerEnsslin}].
}
\label{Setup2}
\end{center}
\end{figure}

We consider a lateral quantum dot~\cite{KAT}
formed by depleting the 2DEG
via a set of metallic gates that allow the quantum dot
position ${\bm r}_0$ to be changed at will to
distances comparable to the spin-orbit length
in the 2DEG.
Such rolling quantum dots can be defined using,
for example, a set of gates shown in Fig.~\ref{Setup1}.
Two layers of finger-like gates (separated by an insulator) 
form a grid, which construct the dot confining potential
at virtually any position under the grid by simultaneously pulsing  
several gates.
A relatively different setup is shown in Fig.~\ref{Setup2}, which
makes use of a quantum ring and allows the quantum dot to be moved along 
a circular trajectory.

The electron motion in the plane of the 2DEG, and in the presence
of a time-dependent dot confining potential, is governed by the Hamiltonian
\begin{equation}
H(t)=H_d(t)+H_Z+H_{SO},
\label{HoftHdoftHZHSO}
\end{equation}
where $H_d(t)$ describes the moving dot with one electron,
\begin{equation}
H_d(t)=\frac{p^2}{2m_e}+U\left({\bm r}-{\bm r}_0(t)\right),
\label{Hdofthere}
\end{equation}
with ${\bm p}=-i\hbar\partial/\partial{\bm r}+(e/c){\bm A}({\bm r})$ and ${\bm r}=(x,y)$
being the electron momentum and coordinates, respectively.
For the vector potential ${\bm A}({\bm r})$, we choose the cylindric gauge,
${\bm A}({\bm r})=B_z\left(-y/2,x/2,0\right)$, where $B_z$ is the
component of the magnetic field normal to the 2DEG plane.
The dot confinement potential $U({\bm r},t)=U\left({\bm r}-{\bm r}_0(t)\right)$ 
changes adiabatically with respect to the size-quantization energy in the dot and, 
thus, no transitions between orbital levels occur.
For simplicity, we also assume that the shape of the dot confinement 
does not change in time, while the dot is moved along its trajectory ${\bm r}_0(t)$.
In Eq.~(\ref{HoftHdoftHZHSO}),
the Zeeman interaction reads
$H_Z=\frac{1}{2}{\bm E}_Z\cdot\mbox{\boldmath $\sigma$}$, 
with $|{\bm E}_Z |$ being the Zeeman energy and
 $\mbox{\boldmath $\sigma$}=(\sigma_x,\sigma_y,\sigma_z)$
the Pauli matrices.
We note that the quantization axis is generally not along the magnetic field and one has
 $E_{Zi}=\mu_{\rm B}{\rm g}_{ij}B_j$, where $\mu_{\rm B}$ 
is the Bohr magneton, ${\rm g}_{ij}$ is the ${\rm g}$-factor tensor in the 2DEG,
and ${\bm B}$ is the magnetic field.
The last term in Eq. (\ref{HoftHdoftHZHSO}), $H_{SO}$, denotes the spin-orbit 
interaction which has the 
following general form
$H_{SO}=\frac{1}{2}{\bm h}({\bm p})\cdot\mbox{\boldmath $\sigma$}$, 
where ${\bm h}({\bm p})=-{\bm h}(-{\bm p})$ is an odd-power polynomial in ${\bm p}$.
In GaAs 2DEG with the $[001]$ growth direction, for example,
the leading order (lowest power in ${\bm p}$) spin-orbit interaction terms read
\begin{equation}
H_{SO}=\alpha(p_x\sigma_y-p_y\sigma_x)+\beta(-p_x\sigma_x+p_y\sigma_y),
\label{HSO}
\end{equation}
where $\alpha$ and $\beta$ are the Rashba and Dresselhaus coupling constants,
respectively \cite{Rashba,Dress} .

Considering first a stationary quantum dot potential $U({\bm r})$, 
we set ${\bm r}_0(t)\to 0$ in Eq.~(\ref{Hdofthere}) and
denote the time-independent Hamiltonian by $H$ with the following eigenvalue equation, 
$H|\psi_{ns}\rangle=E_{ns}|\psi_{ns}\rangle$,
where $n=0,1,2\dots$ and $s=\pm 1/2$
are the orbital and spin quantum numbers, respectively.

The eigenstates of $H$ are related to the eigenstates of $H_d+H_Z$
via a unitary transformation (see  Appendix)
\begin{equation}
\left|\psi_{ns}\right\rangle=e^{-S}\left|\psi_{n}\right\rangle\left|\chi_s\right\rangle,
\label{Psinsexppsinchis}
\end{equation}
where $|\psi_{n}\rangle$ and  $\left|\chi_s\right\rangle$ are obtained by solving 
$H_d|\psi_{n}\rangle=E_n|\psi_{n}\rangle$ 
and $H_Z\left|\chi_s\right\rangle=sE_Z\left|\chi_s\right\rangle$, respectively.

We are mainly interested here in spin dynamics in the {\it absence}
 of the applied magnetic field, therefore, we set ${\bm A}({\bm r})=0$,
and focus on the lowest-in-energy subspace $n=0$ (for the discussion of the spin dynamics in the presence of an applied magnetic field, see  Appendix). 
The suitable qubit is then defined as 
$\left|\uparrow\right\rangle=|\psi_{0,1/2}\rangle$\; and\;
$\left|\downarrow\right\rangle=|\psi_{0,-1/2}\rangle$.
In the absence of magnetic fields, the quantization axis can, therefore, be chosen arbitrarily. However, once it is chosen, all subsequent spin rotations are then with respect to this axis.

Given a linear in momentum spin-orbit interaction in Eq. (\ref{HSO}),
and a symmetric confining potential
$U({\bm r})=U(-{\bm r})$, we have explicitly found the generator of the rotation for
the Kramers doublet (see  Appendix), as one moves the dot along
a given path
\begin{eqnarray}
\Delta = {\mathbbm {1}}-i\mbox{\boldmath $\sigma$}\cdot {\bm{\lambda}}_{SO}^{-1}\cdot\delta{\bm r}_0,
\label{trans1plFFBeq0operator2}\;\;\;\;\;\;\;\;\;\;\;\;\;\;\;\;\; \\
{\bm{\lambda}}_{SO}^{-1} \equiv \left(
\begin{array}{cc}
0&1/\lambda_-\\
1/\lambda_+&0
\end{array}
\right), \;\;\;\;
\lambda_{\pm}= \frac {\hbar}{m_e (\beta \pm \alpha)}, \label{lambdatensor}
\end{eqnarray}
where $\mbox{\boldmath$\sigma$}$ are Pauli matrices acting on the
Kramers doublet states and
$S=i\mbox{\boldmath $\sigma$}\cdot{\bm{\lambda}}_{SO}^{-1}\cdot{\bm r}$ \cite{GBL} .
According to Eq.~(\ref{trans1plFFBeq0operator2}),
the electron state, in the space of a Kramers doublet,
is rotated during the displacement by an angle $\sim \delta r_0/\lambda_{SO}$.
This interpretation of the spin-orbit interaction effect is identical to
the standard interpretation given to semiclassical electrons.
In the latter,
the electron with the momentum $p$ travels a distance
$l=pt/m_e$ during a time $t$
and changes its spin by an angle proportional to $l/\lambda_{SO}$.
This coincidence is not accidental, because in the semiclassical picture one also
assumes that the electron moves in a wave packet of an extension that is
much smaller than $\lambda_{SO}$.
The speed at which the electron moves is unimportant at $B=0$, because the path 
${\bm r}_0(t)$ is the only information that determines the spin rotation.
In particular, if the electron travels along some path forward and then returns
the same way, but not necessarily at the same speed,
then the initial and final spin states coincide.
\begin{figure}
\begin{center}
\includegraphics[angle=0,width=0.4\textwidth]{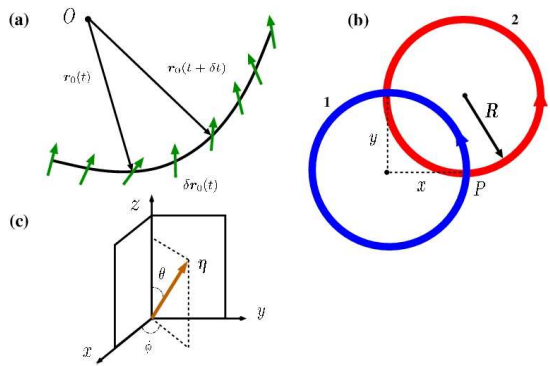}
\caption{\small
(Color online) {\bf{(a)}} Trajectory of the quantum dot center 
${\bm r}_0(t)$ and the evolution of the spin 
state due to displacement.
The spin state changes when going from ${\bm r}_0$ to
${\bm r}_0+\delta{\bm r}_0$, due to the infinitesimal transformation (\ref{trans1plFF}).
Since the directions of $\delta{\bm r}_0$ can be different in different parts of
the curve, the infinitesimal transformations do not commute with each other
and have to be ordered along the path of integration.
{\bf{(b)}} Two oriented circles with the same radii $R$ but different centers
(with their corresponding holonomic rotations at $P$) show the typical paths for the
quantum dot on the 2DEG. 
{\bf{(c)}} Moving the dot along the circles will rotate the spin of the electron
around the $\bm {\eta}$ axis.  For instance, blue (1) and red (2) circles in (b) correspond 
to $\phi=0$ and $\phi=-\frac{\pi}{2}$, respectively.
}
\label{trajectoryfigure}
\end{center}
\end{figure}

We note that, for the linear in momentum $H_{SO}$, 
the tensor ${\bm{\lambda}}_{SO}^{-1}$ in Eq. (\ref{lambdatensor})
is independent of the electron orbital state $|\psi_{n}\rangle$, which means 
Eq.~(\ref{trans1plFFBeq0operator2})
is valid for all symmetric wave packets at the zeroth order of $\lambda_d/\lambda_{SO}$.
Therefore, we can consider also a point-like electron, for which
the orbital wave function reads
$\psi_{{\bm r}_0}({\bm r})=\delta({\bm r}-{\bm r}_0)$.
Integrating Eq.~(\ref{trans1plFFBeq0operator2}) over an arbitrary path we obtain
an exact expression for this case,
\begin{equation}
\psi_{{\bm r}_0s}({\bm r},\sigma)=
e^{-i\int{\bm \sigma}\cdot{\bm{\lambda}}_{SO}^{-1}\cdot d{\bm r}_0}
\delta({\bm r}-{\bm r}_0)\chi_{s}(\sigma),
\label{exactpsir0sexpdelta}
\end{equation}
where the exponent is ordered (to the left) along the path of integration.
To simplify notations, we use $\int d{\bm r}_0$
to denote the contour integral $\int_0^{{\bm r}_0}d{\bm r}'$ and
agree that any exponent of an integral to be ordered along the path 
of integration.
The radius vector ${\bm r}_0(t)$ gives us the path where we choose the 
beginning of the path at ${\bm r}_0=0$, and denote
the running (present) point of the path by ${\bm r}_0$.
Eq.~ (\ref{exactpsir0sexpdelta}) determines how the spin of an electron
is transformed (at $B=0$) as the electron is moved along an arbitrary path.
The difference between the spin and the Kramers doublet disappears
here, since for  point-like electrons we can take $S\to 0$.
However, 
the transformation rule in Eq.~(\ref{exactpsir0sexpdelta}) arises
from the fact that $\partial S/\partial {\bm r}$ remains constant
while taking $S\to 0$.
Moreover, it holds exactly for
the linear in momentum spin-orbit interaction, because
${\bm{\lambda_{SO}}}^{-1}$ is independent of the orbital state.
For the $p^3$ terms, $1/\lambda_{SO}$ is proportional to
the electron energy and Eq.~(\ref{exactpsir0sexpdelta})
can be written only if $H_{SO}$ is first linearized around 
a given energy.

We developed a code to calculate numerically the spin dynamics for an arbitrary path, 
however as an example, we consider a point like quantum dot (strong confining potential) 
moving around a circle with radius $R$ (see Fig. \ref{Setup2}).
Using Eq.~(\ref{trans1plFFBeq0operator2}), the dynamical equation for the 
electron spin is then given by
\begin{eqnarray}
\frac{d}{d\phi}\chi_s(\phi)= i {\cal M}_{DR} \chi_s(\phi), \label{Spindynamics} \\
{\cal M}_{DR} \equiv R (\frac{\sin \phi}{\lambda_+}\sigma_y -
 \frac{\cos \phi}{\lambda_-}\sigma_x ),
\label{Spinevolution}
\end{eqnarray}
where $\chi_s(\phi)=(v_1,v_2)$ is the spinor, $\phi$ is the angle between the 
starting point vector ${\bm r}_0(0)$ and the $x$ axis, and ${\cal M}_{DR}$ is a 
Hermitian matrix due to both 
Dresselhaus (${\cal M}_D$)
and Rashba (${\cal M}_R$) spin orbit interaction (e.g. for the only Rashba interaction 
we have ${\cal M}_{DR}={\cal M}_{R}=-{\cal M}_{D}^*$). In addition to trivial solutions for a linear path in Eq.~(\ref{exactpsir0sexpdelta}), there are also simple analytical solutions for any elliptic or hyperbolic path ($x^2/a^2 \pm y^2/b^2 =1$, where $a$ and $b$ are conic section parameters), provided
$a/ \lambda_+=b/ \lambda_-$.
 We show here the solution for
a $\phi=2\pi$ rotation, counterclockwise along the blue (1) circle in Fig.~\ref{trajectoryfigure},
for Dresselhaus only where ${\cal M}_{DR}={\cal M}_{D}$ and $\lambda_+=\lambda_- \equiv \lambda R$
\begin{eqnarray}
\chi_s(2\pi) &=& \exp(-\frac{i}{2}{\bm \eta} \cdot {\bm \sigma}) \chi_s(0),\label{holonomy1}\\
{\bm \eta} &=& 2 \pi (1-\frac{1}{\epsilon})(\frac{2}{\lambda},0,1),
\label{holonomy2}\\
\epsilon &=& \sqrt{1+4/ \lambda^2}. \nonumber
\end{eqnarray} 
Mathematically speaking, transformation~(\ref{holonomy1}) is an element of the holonomy 
group of point $P$, generated by the non-Abelian vector potential 
\cite{Zanardi, Nakahara} where
this non-Abelian feature is a direct consequence of the Kramers degenerate doublets. Therefore,
in the absence of any time reversal symmetry breaking interactions (like e.g. an applied magnetic field or magnetic impurities),
the Kramers doublets are robust degenerate states and we are able to use the non-Abelian feature
of the effective gauge potential to manipulate the spin.

\section{Single-qubit Rotations: Hadamard and Phase Gates}

The  transformation (\ref{holonomy1}) is nothing but a rotation along the vector 
${\bm \eta}$ at point $P$ which is in the $x$-$z$ plane and makes an angle 
$\theta=\arctan (\frac{2}{\lambda})$ with the $z$ axis,
see Fig.~\ref{trajectoryfigure}.
 Therefore, for large values of $\lambda$
(small circles), $\theta \approx 0$ and one would able to realize arbitrary rotations around $z$ axis 
(phase gate) with reasonable precision. Moreover, moving counterclockwise along 
the red (2) circle in Fig.~\ref{trajectoryfigure} leads to 
the same result as in the blue (2) circle, but now the rotation takes place 
in the $y$-$z$ plane. 
Therefore, depending on the orientation of the circles and their corresponding radii, 
we can, in principal, achieve all kinds of rotations around the Bloch sphere.

To be more specific, we show how to generate the Hadamard gate by rotations around two 
non-orthogonal axes (in our case $z$ and
${\bm \eta}$ directions) \cite{Guido}  and, for convenience, we only consider the Dresselhaus term ($\alpha =0$). 
As shown in Eq. (\ref{holonomy1}),  circles with different radii and/or
orientations lead to different rotations. In particular, if we go counterclockwise along a full circle
from point $P$ which makes an angle $\phi$ with $x$ axis (see Fig. \ref{trajectoryfigure}), 
the electron spin will transform as follows  
\begin{eqnarray}
U_{11} &=& -\cos{\pi \epsilon}+
\frac{i}{\epsilon} \sin {\pi \epsilon},\;\;\;\;\;\; \nonumber \\
U_{12} &=& \frac{2 i e^{i\phi}}{\lambda \epsilon} \sin {\pi \epsilon}, \nonumber \\
U_{11} &=& U_{22}^*,\;\;\;\;\;\;\;\;\;\;\;\;\;\;\; U_{12} = -U_{21}^*, \nonumber
\end{eqnarray}
where  $U_{ij}$ is the unitary transformation which acts on the initial spin state.
Geometrically, $U$ is the matrix corresponding to a rotation around $\bm {\eta}$ axis which lies in
the same plane as $z$ axis and $P$ (see Fig. \ref{trajectoryfigure}.c).

Hadamard gate can be achieved (up to a global phase) by a clockwise $\frac{\pi}{2}$ rotation 
around $y$ axis followed by a counterclockwise $\pi$ rotation around $z$ axis,
\begin{equation}
H  = \frac{1}{\sqrt{2}} \left(
\begin{array}{cc}
1&1\\
1&-1
\end{array}
\right) = i\;U_z(\pi)\;U_y(-\frac{\pi}{2}).
\end{equation}
The rotation around $z$ with an arbitrary angle has already been discussed above, therefore, 
we show here how one can implement $U_y(\frac{\pi}{2})$. The main problem is that, 
according to Eqs.~[\ref{holonomy1},\ref{holonomy2}], the magnitude and the direction of the vector 
$\bm {\eta}$ are not independent variables (both are functions of the variable  $\lambda$).
Moreover, the vector $\bm{\eta}$ 
is not in general orthogonal to $z$ axis. To obtain an arbitrary rotation around, say, $y$ axis, we 
need to perform $3$ rotations (two around $\bm{\eta}$ and one around $z$ axis). 
Specifically, we want to know for which values of $\lambda$,  $\pm\frac{\pi}{2}$ rotations around  
$y$ axis can be achieved. One can show that \cite{Guido}
\begin{eqnarray}
U_y (\gamma) &=& U_{\eta} (\theta) U_z (\phi)  U_{\eta} (\theta),  \label{U}\\
\cos\eta &=& \frac{\cos^2\theta\sin^2\frac{\gamma}{2} \pm \cos\frac{\gamma}{2}
\sqrt{1-\cot^2\theta\sin^2\frac{\gamma}{2}}}
{\cos^2\theta\sin^2\frac{\gamma}{2}-1},\;\;\;\;\;\;\; \label{y-rotation} \nonumber \\
\tan {(\frac{\phi}{2})} &=& -\frac{\sin\eta\cos\theta}{\sin^ 2{\theta}+
\cos{\eta}\cos^2{\theta}}, 
\;\;\;\;\;\;\;\;\;\;\; \eta = |\bm{\eta}|, \nonumber 
\end{eqnarray}
where for our purpose, $U_y(\frac{\pi}{2})$, we need to evaluate Eq.~(\ref{U}) 
at $\gamma =\pi/2$. 
Obviously, there are an infinite number of solutions corresponding to different values of $\lambda$.
 Therefore, the $\pi/2$ rotation around $y$ axis, and consequently 
the Hadamard gate, is achievable in our scheme. Together with the phase gate 
(arbitrary rotation around $z$ axis),
 all single qubit operations can be realized.

We note that the quantum fluctuations of the driving field will lead
to the decoherence of the Kramers doublets even in the absence of
the applied magnetic field, however, this rate saturates at the zero field strength \cite{Pablo}.
For typical lateral GaAs quantum dots with the size $\lambda_d \sim 50$ nm which
 corresponds to the orbital
quantization $\omega_0 \sim 1$ meV, the estimated decoherence rate 
$\Gamma \sim \mu s^{-1}$ \cite{Pablo}.
 Obviously,  to implement efficient single qubit gates by using a ring of tunnel-coupled 
quantum dots,
the pulsing and the total travel time of the quantum dot should
be then much smaller than the decoherence time $\Gamma^{-1}$.
On the other hand, the adiabaticity criterion puts a limit on 
the velocity $\bm v$ of the moving quantum dot in order to keep the electron
in its ground state doublet.
Therefore, one needs to satisfy $ \Gamma \lambda_{SO}   \ll | \bm v | \ll \lambda_d \omega_0$ 
at any moment
in time for a quantum dot which is displaced on the scale of $\lambda_{SO}$.
Recent experiments on tunnel-coupled quantum dots show the ability to transfer
the electron wave function over few hundreds of nm (from one dot to the other) in 1 ns \cite{KAT}.
We observe that, for instance, for the spin-orbit length $\lambda_{SO} \sim 3$ $\mu$m, 
this pulsing time is perfectly within 
the range of the above mentioned condition and our scheme is, therefore, experimentally
feasible.

\begin{figure}
\begin{center}
\includegraphics[angle=0,width=0.4\textwidth]{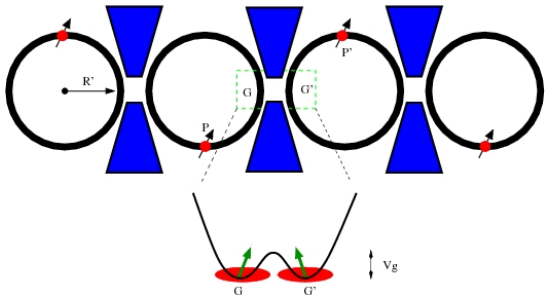}
\caption{\small
(Color online)
Schematic of the two-qubit rotation setup. The confined electrons are brought together 
at the intersection of the circular paths (with their potential profile shown in the inset at this point),
 where they interact via Heisenberg exchange interaction. The (holonomic) single-qubit operations
 are done at position $P$ ($P'$) while the (exchange) two-qubit gates are performed at 
position $G$ ($G'$).}
\label{2qubit}
\end{center}
\end{figure}

\section{Two-qubit Gates, Read Out, and Quantum Computation}
For quantum computing, in addition to single qubit rotations, two qubit operations are needed 
as well (e.g. CNOT gate). Here, we propose a scalable setup to perform quantum computation 
on arbitrary $N$ qubits (see. Fig. \ref{2qubit}). To this end, we move two quantum dots 
around a circular path and bring them close to each other, e.g. from the 
position $P$ ($P'$) to $G$ ($G'$), as shown in Fig. \ref{2qubit}. The top gates are introduced to
control over the wave functions of the confined electrons at the touching points of the circles.
 By lowering the potential barrier ($V_g$) between two quantum quantum dots, 
the residing electron spins can couple to
 each other (due to the overlap of their orbital wave functions) via the Heisenberg exchange 
interaction \cite{LD,BLD}. It has already been shown 
that by electrically engineering the gate potentials, one can generate the SWAP gate, and eventually the CNOT gate, between two spins \cite{LD}.
 However, this additional step, i.e. moving two 
quantum dots towards each other to perform two-qubit gates, leads to a spin rotation of each electron,
 see Eqs. (\ref{Spindynamics},\ref{Spinevolution}) , and 
therefore this partial spin dynamics should also be taken into account. 
For practical purposes, we assume that the radius of the (holonomic) single-qubit gates ($R$)
is smaller than  the radius of the two-qubit circles ($R'$),
 to avoid spatial overlap of different local single-qubit operations,
 see  Figs. \ref{trajectoryfigure}.b,\ref{2qubit}.
The read-out part of the scheme is accomplished  by applying an external magnetic field, using different techniques such as spin-to-charge conversion
\cite{KAT}.

 As an example, we observe that our scheme is able to produce the so called  
{\it cluster states} on $N$ qubits \cite{br}. 
The peculiar properties of cluster states made them a suitable candidate
 for realizing quantum computers in quantum optical and solid state systems 
\cite{br,rauss,BLcluster,MQC-Review}. 
One of the main advantages of one-way quantum computation is that
these set of entangled states (cluster states) are produced once, and then quantum 
computation is done by local (adaptive) measurements of the qubits. 
Therefore, there is no need to perform two-qubit gates during the information processing.
In Fig. \ref{2qubit}, we showed a one dimensional linear chain of qubits, however, 
this scheme can be easily generalized to higher dimensional lattices, and in principal be
 used to generate cluster states and perform
{\it holonomic one-way} quantum computing in solid state environments.

\section*{Acknowledgments}

We would like to thank Andreas Aste and Guido Burkard for useful discussions.
This work was supported by the Swiss NSF, NCCR Basel, EU RTN "Spintronics", and U.S. DARPA.

\appendix*

\section{The Non-Abelian Spin Rotation Generator}

The transformation matrix $S=-S^\dagger$ in Eq. (\ref{Psinsexppsinchis}) can be evaluated 
by perturbation theory in $H_{SO}$ or by diagonalizing the Hamiltonian $H$ for
a specific potential $U({\bm r})$.
At the leading order in $H_{SO}$, $S$ satisfies the operator equation
\begin{equation} 
[H_d+H_Z,S]=H_{SO},
\end{equation} 
whereas the energy levels of $H$ coincide, at this order, with the energy levels of $H_d+H_Z$.
A detailed analysis of the transformation in Eq.~(\ref{Psinsexppsinchis})
is given in Refs.~[\onlinecite{GBL,BGL}].

Next we consider the Schr\"{o}dinger equation,
\begin{equation}
i\hbar\frac{\partial\Psi({\bm r},t)}{\partial t}=H(t)\Psi({\bm r},t),
\label{schroedinger}
\end{equation}
in the presence of a time-dependent displacement-vector ${\bm r}_0(t)$.
At each moment in time, the Hamiltonian $H(t)$ has an 
instantaneous basis of states, which we denote by $|\Phi_{ns{\bm r}_0}\rangle$,
where the index ${\bm r}_0$ indicates that the dot is centered at ${\bm r}_0$.
Obviously, the states $|\Phi_{ns{\bm r}_0}\rangle$
can be obtained from Eq.~(\ref{Psinsexppsinchis})
by means of a displacement by the vector $-{\bm r}_0$.
In the presence of a magnetic field, 
the instantaneous eigenstates read
\begin{equation}
\Phi_{ns{\bm r}_0}({\bm r})=e^{(ie/\hbar c)f\left({\bm r},{\bm r}_0\right) 
}T_{-{\bm r}_0}\psi_{ns}\left({\bm r}\right),
\label{Phinsr0def}
\end{equation}
where $f\left({\bm r},{\bm r}_0\right)$ is a gauge function 
satisfying the equation
\begin{equation}
-\frac{\partial f\left({\bm r},{\bm r}_0\right)}{\partial {\bm r}}=
{\bm A}({\bm r}-{\bm r}_0)-{\bm A}({\bm r}).
\end{equation}
In Eq.~(\ref{Phinsr0def}), 
$T_{{\bm a}}=\exp({\bm a}\cdot\partial/\partial{\bm r})$ denotes the translation
operator by a vector ${\bm a}$ and
for the cylindric gauge ${\bm A}({\bm r})=B_z\left(-y/2,x/2,0\right)$, 
we can choose
$f\left({\bm r},{\bm r}_0\right)={\bm r}_0\cdot {\bm A}({\bm r})$.

The solutions of the Schr\"{o}dinger equation (\ref{schroedinger}) can be looked for
in terms of the instantaneous basis in Eq. (\ref{Phinsr0def}),
\begin{equation}
\Psi({\bm r},t)=\sum_{ns}a_{ns}(t)\Phi_{ns{\bm r}_0(t)}\left({\bm r},t\right),
\label{Psinrtsumasphi}
\end{equation}
where the coefficients $a_{ns}(t)$ satisfy the normalization condition
$\sum_{ns}|a_{ns}(t)|^2=1$ and we have used the notation
$\Phi_{ns{\bm r}_0(t)}\left({\bm r},t\right)=
\exp\left(-iE_{ns}t/\hbar\right)\Phi_{ns{\bm r}_0(t)}\left({\bm r}\right)$
for the Schr\"odinger picture of Eq.~(\ref{Phinsr0def}).
By substituting Eq.~(\ref{Psinrtsumasphi}) into Eq.~(\ref{schroedinger})
 we obtain a set of equations for $a_{ns}(t)$
\begin{equation}
\frac{da_{ns}}{dt}= 
\frac{i}{\hbar}\sum_{n's'}
{\bm v}_0(t)\cdot\left\langle\psi_{ns}\right|\tilde{\bm p}(t)
\left|\psi_{n's'}\right\rangle 
e^{i\omega_{nsn's'}t} a_{n's'},
\label{coefans}
\end{equation}
where ${\bm v}_0(t)=d{\bm r}_0(t)/dt$ is the velocity of the slipping dot and
$\omega_{nsn's'}=\left(E_{ns}-E_{n's'}\right)/\hbar$.
The quantity $\tilde{\bm p}(t)$
depends on $t$ only through ${\bm r}_0(t)$ and is defined as follows
\begin{equation}
\tilde{\bm p}=-i\hbar\frac{\partial}{\partial {\bm r}}
-\frac{e}{c}T_{{\bm r}_0}
\frac{\partial f({\bm r},{\bm r}_0)}{\partial {\bm r}_0}
T_{-{\bm r}_0}.
\end{equation}
For our choice of gauge, i.e. cylindrical gauge, we obtain
$\tilde{\bm p}=-i\hbar\partial/\partial{\bm r} -(e/c){\bm A}({\bm r}+{\bm r}_0)$.
Note that the choice of $f\left({\bm r},{\bm r}_0\right)$ is not unique;
In general, one can also include 
terms of the form $g_0({\bm r}_0)+sg_3({\bm r}_0)\left|\psi_{ns}\rangle\langle\psi_{ns}\right|$
and, at $B=0$, additionally terms of the form
$g_1({\bm r}_0)\left|\psi_{ns}\rangle\langle\psi_{n,-s}\right|+
isg_2({\bm r}_0)\left|\psi_{ns}\rangle\langle\psi_{n,-s}\right|$,
where $g_j({\bm r}_0)$ are arbitrary real functions of ${\bm r}_0$.

Next we consider a specific situation for which we can further simplify
Eq.~(\ref{coefans}).

We can further define a resting qubit at a position ${\bm r}_0$
using the transformation in Eq.~(\ref{Phinsr0def}).
Let the quantum dot be driven along a trajectory ${\bm r}_0(t)$
between two points ${\bm r}_A=0$ and ${\bm r}_B$ during
a time interval $T$, such that
\begin{eqnarray}
&{\bm r}_0(0)={\bm r}_A,
\;\;\;&\;\;\;
{\bm r}_0(T)={\bm r}_B,\\
&{\bm v}_0(0)=0,
\;\;\;&\;\;\;
{\bm v}_0(T)=0.
\end{eqnarray}
The probability for the qubit to leak out
of its subspace by the end of the pulse is given by
\begin{equation}
P_{\rm leak}=
\sum_{n \neq\, 0 \atop s = \pm 1/2}
\left|a_{ns}(T)\right|^2.
\end{equation}
The coefficients $a_{ns}(T)$ can, therefore, be found by solving Eq.~(\ref{coefans})
with the initial condition $\sum_{s}\left|a_{0s}(0)\right|^2=1$.
At the leading order in the driving, we have
\begin{equation}
a_{ns}(T)\simeq
\frac{i}{\hbar}
\int_0^T dt \;
{\bm v}_0(t)\cdot\left\langle\psi_{ns}\right|\tilde{\bm p}(t)
\left|\sigma\right\rangle
e^{i\omega_{ns0\sigma}t} ,
\label{ansperturbtheory}
\end{equation}
where $\left|\sigma\right\rangle=\left|\psi_{0\sigma}\right\rangle$ 
denotes the qubit state at $t=0$.
Since the matrix elements 
$\left\langle\psi_{ns}\right|\tilde{\bm p}(t)\left|\sigma\right\rangle$
do not depend on time for $n\neq 0$, 
the coefficients $a_{ns}(T)$ in
Eq.~(\ref{ansperturbtheory}) are, thus, proportional to the Fourier
transform of the quantum dot velocity ${\bm v}_0(t)$ evaluated  
at the orbital transition frequency $\omega_{ns0\sigma}\simeq\omega_0$.
It is, therefore, sufficient to devise pulses of ${\bm v}_0(t)$
with the spectral weight below the orbital frequency $\omega_0$ in order to
avoid leakage from the qubit subspace.

It is convenient to have an adiabaticity criterion based on the
differential properties of ${\bm v}_0(t)$.
We note that Eq.~(\ref{coefans})
can be rewritten in terms of the new unknowns
$\tilde{a}_{ns}=a_{ns}\exp\left(-itE_{ns}/\hbar\right)$ as follows
\begin{eqnarray}
\frac{d\tilde{a}_{ns}}{dt}&=&
-\frac{i}{\hbar}\sum_{n's'}
{\cal H}_{nsn's'}(t)
\tilde{a}_{n's'},
\label{coeftildeans}\\
{\cal H}_{nsn's'}(t)&=&E_{ns}\delta_{ns,n's'}-
{\bm v}_0(t)\cdot\left\langle\psi_{ns}\right|\tilde{\bm p}(t)
\left|\psi_{n's'}\right\rangle.\;\;\;\;\;\;\;\;\;\;\;
\label{calHnsns}
\end{eqnarray}
One can identify Eq.~(\ref{calHnsns}) with
${\cal H}(t)=H(t)-i\hbar\partial/\partial t$, 
expressed in the time-depended basis (\ref{Phinsr0def}).
For the case of $B=0$, 
Eq.~(\ref{calHnsns}) has been previously obtained 
in Ref.~[\onlinecite{Pablo}].
The virtue of Eq.~(\ref{calHnsns}) is that
${\cal H}_{nsn's'}(t)$ depend on time only through
a perturbation $\propto v_0(t)$, which vanishes at $t=0,T$.
Applying the adiabaticity criterion  to 
the Hamiltonian in Eq.~(\ref{calHnsns}) for
the orbital transitions out of the qubit subspace,
we obtain that the condition
\begin{equation}
\left | \frac{d{\bm v}_0}{dt}\cdot\left\langle\psi_{ns}\right|\tilde{\bm p}\left|
\sigma\right\rangle \right |
\; \ll \; \hbar\omega_0^2
\end{equation}
must be satisfied at any moment in time in order for the pulse to be adiabatic.

If ${\bm r}_0(t)$ changes adiabatically with respect also to the
Zeeman energy $E_Z$, then $|a_s(t)|$ is independent of time,
i.e. the qubit follows adiabatically the change of its basis states.
In the opposite case, when $B=0$,
the states $\left|\psi_{ns}\right\rangle$ in Eq.~(\ref{Psinsexppsinchis})
are degenerate with respect to the spin index $s$ 
to all orders of $H_{SO}$, due to the Kramers theorem.
In this case, the change of the instantaneous basis can be 
interpreted as a unitary operation on the qubit.
In order to tell what is the qubit instantaneous basis at $B=0$,  
one has, in principle, to consider a finite $B$ and
follow the energy levels of the quantum dot in the limit of $B\to 0$.
Here, it is important to note that the spin-orbit interaction gives rise
to an anisotropic Zeeman interaction at the second order of $H_{SO}$ 
\cite{GBL}.
As a result, the spin quantization axis and the magnetic field
are not necessarily aligned with each other.
 To avoid the need of state finding, we denote
 $\langle ns|e^{-S}|n,-s\rangle$ by $\alpha_{ns}$
 and remark that $\alpha_{ns}={\cal O}\left(H_{SO}^2\right)$.

Returning now to Eq.~(\ref{Psinrtsumasphi}),
we consider an infinitesimal displacement of the quantum dot
in the ($x$,$y$)-plane by $\delta{\bm r}_0$ and derive the 
corresponding generators of the qubit transformation under translations.
We encode the qubit into the instantaneous states of the $n$-th orbital level
of the Hamiltonian (\ref{HoftHdoftHZHSO}).

Let ${\bm r}_0(t)={\bm r}_0$ be the position of the quantum dot center
at time $t$ and ${\bm r}_0(t+\delta t)={\bm r}_0+\delta{\bm r}_0$
be the new position at time $t+\delta t$.
The infinitesimal transformation
that takes the state $\Psi(t)$ to a new state $\Psi(t+\delta t)$
is given in Eq.~(\ref{schroedinger}).
Starting from a basis state $\Phi_{ns}(t)$ at
time $t$, we obtain the following state
at time $t+\delta t$,
\begin{equation}
\Psi_{ns}(t+\delta t)=\Phi_{ns}(t)-\frac{i}{\hbar}H(t)\Phi_{ns}(t)\delta t.
\label{psitplusdeltat}
\end{equation}
The overlap of this state with
the basis state $\Phi_{ns'}(t+\delta t)$, 
generates an infinitesimal transformation of the Kramers
doublet, where
from Eq.~(\ref{Phinsr0def}),
we find the basis state at time $t+\delta t$,
\begin{equation}
\Phi_{ns'}(t+\delta t)=\Phi_{ns'}(t)
-\frac{i}{\hbar}\left[\frac{d{\bm r}_0}{dt}\cdot\tilde{\bm p}+H(t)\right]
\Phi_{ns'}(t)\delta t.
\label{phinsprtpldtplplpl}
\end{equation}
Thus, the desired infinitesimal transformation reads
\begin{equation}
\langle \Phi_{ns'}(t+\delta t)|\Psi_{ns}(t+\delta t)\rangle= 
\delta_{s's}+\frac{i}{\hbar}\delta{\bm r}_0\cdot
\langle \Phi_{ns'}(t)|\tilde{\bm p}|\Phi_{ns}(t)\rangle.
\label{trans1plFF}
\end{equation}
For a qubit that is encoded into the instantaneous states
of the $n$-th orbital level 
of the Hamiltonian (\ref{HoftHdoftHZHSO}),
the infinitesimal transformation (\ref{trans1plFF})
can be rewritten as
\begin{equation}
|s(t)\rangle\rightarrow \exp\left(\mbox{\boldmath $\cal G$}\cdot\delta{\bm r}_0\right)|s(t)\rangle.
\label{EvolvQubit}
\end{equation}
Here,  $|s(t)\rangle$ denotes the qubit state at time $t$ and the $2\times 2$ matrices
\begin{equation}
\mbox{\boldmath $\cal G$}_{ss'}=\frac{i}{\hbar}
\langle \psi_{ns}|
\tilde{\bm p}
|\psi_{ns'}\rangle
\label{GenG}
\end{equation}
are the corresponding generators of the transformations
that take place on the qubit under parallel translations of 
the quantum dot on the substrate.
In deriving Eq.~(\ref{GenG}) we made use of our choice of gauge,
see the text below Eq.~(\ref{Phinsr0def}).

It is important to note that, along with spin-orbit interaction-induced 
${\rm SU}(2)$ transformations on the qubit, Eqs.~(\ref{EvolvQubit}) and (\ref{GenG}) 
account also for the Aharonov-Bohm phase due to the orbital magnetic field.
It is, therefore, convenient to subdivide $\mbox{\boldmath $\cal G$}$
into Abelian and non-Abelian parts,
\begin{eqnarray}
\mbox{\boldmath $\cal G$}_{ss'}&=&\mbox{\boldmath $\cal G$}_{ss'}^{a}+\mbox{\boldmath $\cal G$}_{ss'}^{na},\\
\mbox{\boldmath $\cal G$}_{ss'}^{a}&=&\delta_{ss'}\sum_{p}\frac{1}{2}\mbox{\boldmath $\cal G$}_{pp}.
\end{eqnarray} 
For a point-like quantum dot, we 
obtain the Abelian generators
$\mbox{\boldmath $\cal G$}_{ss'}^{a}=(-ie/\hbar c)\delta_{ss'}{\bm A}({\bm r}_0)$,
recovering, thus, the usual expression for the Aharonov-Bohm phase ($e^{i\varphi_{\rm AB}}$)
\begin{equation}
\varphi_{\rm AB}=-\frac{e}{\hbar c}\int_{\cal C}{\bm A}({\bm r}_0)\cdot d{\bm r}_0
\end{equation}
in going around a closed path ${\cal C}$.
Note that it is always possible to sum up independently
the phase due to the Abelian generators, because
$[\mbox{\boldmath $\cal G$}^{a},\mbox{\boldmath $\cal G$}^{na}]=0$.
In what follows, we focus on the non-Abelian generators $\mbox{\boldmath $\cal G$}^{na}$
since they give rise to useful unitary operations on the qubit.

To calculate the matrix elements 
$\langle \Phi_{ns'}(t)|\tilde{\bm p}|\Phi_{ns}(t)\rangle$,
we make use of Eq.~(\ref{Phinsr0def}) and the following property
$e^{-ief/\hbar c}\tilde{\bm p}e^{ief/\hbar c}={\bm p}$,
and obtain that
\begin{equation}
\langle \Phi_{ns'}(t)|\tilde{\bm p}|\Phi_{ns}(t)\rangle=e^{\frac{i}{\hbar}(E_{ns'}-E_{ns})t}
\langle \psi_{ns'}|T_{{\bm r}_0}{\bm p}T_{-{\bm r}_0}|\psi_{ns}\rangle,
\label{resultmatrixPhinsprnselement}
\end{equation}
where $|\psi_{ns}\rangle$ are the states in Eq.~(\ref{Psinsexppsinchis})
and $E_{ns}$ are the energies corresponding to these states.
Obviously, if the Zeeman energy is large, the exponential factor
in Eq.~(\ref{resultmatrixPhinsprnselement}) oscillates rapidly as a function of time
and the transformation in Eq.~(\ref{trans1plFF}) averages out to unity.
For the latter to take place, it is sufficient that
\begin{equation}
\left | \frac{d{\bm r}_0}{dt}\cdot
\frac{\langle \psi_{ns}|T_{{\bm r}_0}{\bm p}T_{-{\bm r}_0}|\psi_{n,-s}\rangle}{E_{ns}-E_{n,-s}} \right |
\; \ll \; 1.
\end{equation}
Estimating further 
$|\langle \psi_{ns}|T_{{\bm r}_0}{\bm p}T_{-{\bm r}_0}|\psi_{n,-s}\rangle|\sim \hbar/\lambda_{SO}$ 
and $E_{ns}-E_{n,-s}\approx E_Z$,
we obtain that the spin rotator is inefficient at small speeds of the dot,
$\hbar \dot{r}_0 \ll E_Z \lambda_{SO}$.

In the absence of magnetic fields, the
transformation in Eq.~(\ref{trans1plFF})
acquires the form
\begin{equation}
\Delta = {\mathbbm {1}} +\delta{\bm r}_0\cdot
\langle \psi_{ns'}|\frac{\partial}{\partial{\bm r}}|\psi_{ns}\rangle.
\label{trans1plFFBeq0}
\end{equation}
Note that Eq.~(\ref{trans1plFFBeq0}) can as well be derived from the infinitesimal 
version of the identity
$|\Phi_{ns}(t)\rangle=T_{\delta{\bm r}_0}T_{-\delta{\bm r}_0}|\Phi_{ns}(t)\rangle=
T_{\delta{\bm r}_0}|\Phi_{ns}(t+\delta t)\rangle$.
Thus, the spin rotation takes place (at least at $B=0$), 
because the confinement defines local Kramers states, which differ from each other 
along the dot trajectory.
An illustration of the dot trajectory is given in Fig.~\ref{trajectoryfigure}.
The radius-vector ${\bm r}_0(t)$ describes a curve as a function of time
and, as the dot is moved along that curve, the local Kramers state changes.
The infinitesimal transformation in Eq.~(\ref{trans1plFFBeq0}) 
[or more generally in Eq.~(\ref{trans1plFF})]
has to be ordered along the path of integration when integrated over $\delta{\bm r}_0$.
This ordering occurs because
the spin matrices in Eq.~(\ref{trans1plFFBeq0}) do not always 
commute with each other at different points of the path
due to generally different directions of $\delta{\bm r}_0$ at these points.

Further, it is convenient to refer to the Kramers doublets $|\Phi_{ns}(t)\rangle$ as
to spin states that are locally defined at each point of the dot trajectory.
Mathematically, we perform a mapping given by the following canonical transformation,
\begin{equation}
|\Phi_{ns,{\bm r}_0}\rangle=e^{ief/\hbar c}T_{-{\bm r}_0}e^{-S}|\psi_n\rangle|\chi_s\rangle,
\end{equation}
which is obtained by substituting Eq.~(\ref{Psinsexppsinchis}) into Eq.~(\ref{Phinsr0def})
and omitting the free evolution factor $e^{-(i/\hbar)E_{ns}t}$ from $\psi_{ns}({\bm r},t)$.
For a given quantum number $n$, this transformation, obviously, maps the Kramers doublet 
at position ${\bm r}_0$ onto a spin-$1/2$ space: $|\chi_s\rangle$, $(s=\pm 1/2)$.
Equation (\ref{trans1plFFBeq0}) can then be rewritten in an operator form
\begin{equation}
\Delta = {\mathbbm {1}}+\delta{\bm r}_0\cdot
\langle \psi_{n}|e^S\frac{\partial}{\partial{\bm r}}e^{-S}|\psi_{n}\rangle,
\label{transformationdr0esemsoper}
\end{equation}
where $S$ contains Pauli matrices, which should now be regarded as effective
operators in the Hilbert space of a local
Kramers doublet $(n,{\bm r}_0)$.

For a small quantum dot the transformation matrix $S$ is  small, 
because $\lambda_d\ll\lambda_{SO}$ \cite{BGL}.
In this case, one can expand the transformation to the first order, $e^{\pm S}\approx 1\pm S$.
Then, Eq.~(\ref{transformationdr0esemsoper}) acquires the form ($B=0$)
\begin{equation}
\Delta = {\mathbbm {1}}-\delta{\bm r}_0\cdot \langle \psi_{n}|\frac{\partial S}
{\partial{\bm r}} |\psi_{n}\rangle,
\label{trans1plFFBeq0operator}
\end{equation}
where we used $\langle \psi_{n}|\partial/\partial{\bm r}|\psi_{n}\rangle=0$.

\end{document}